\documentclass[letterpaper,12pt]{article}
\usepackage{amsmath,amsthm,amsfonts,amssymb,times,graphics,graphicx,epsfig,subfigure,array,natbib,hyperref,setspace,fancyhdr,enumerate,lineno}
\usepackage[hmargin=1in,vmargin=1in]{geometry}
\linenumbers
\title{The Importance of Prior Choice in Model Selection: a Density Dependence Example}
\author{James D. Lawrence, Robert B. Gramacy,\\Len Thomas and Stephen T. Buckland}
\begin{document}
\begin{spacing}{1.9}
\maketitle
\doublespacing
\begin{abstract}
We perform a Bayesian analysis on abundance data for ten species of North American duck, using the results to investigate the evidence in favour of biologically motivated hypotheses about the causes and mechanisms of density dependence in these species. We explore the capabilities of our methods to detect density dependent effects, both by simulation and through analyzes of real data. The effect of the prior choice on predictive accuracy is also examined. We conclude that our priors, which are motivated by considering the dynamics of the system of interest,
offer clear advances over the priors used by previous authors for the duck data sets.\par
We use this analysis as a motivating example to demonstrate the importance of careful parameter prior selection if we are to perform a balanced model selection procedure. We also present some simple guidelines that can be followed in a wide variety of modelling frameworks where vague parameter prior choice is not a viable option. These will produce parameter priors that not only greatly reduce bias in selecting certain models, but improve the predictive ability of the resulting model-averaged predictor.
\end{abstract}
\section{Introduction}
Density dependence within a species is usually the primary means of numerical self-regulation, the mechanism by which a species can maintain a steady population trajectory in an environment that produces unexpected events of both beneficial and harmful natures.~\cite{turchin95}, in a synthesis of several other sources, states that density dependence is necessary for a regulated population. That is, a population without it is almost certain to be numerically unstable, with an undefined carrying capacity.\par
It is important to discern the magnitude of density dependence a species exhibits, as well as the time lag over which it operates. Knowledge of a species' likely response to natural as well as synthetic shocks will assist in effective species management. Statistically this is a challenging problem which does not usually admit closed-form mathematical analysis.\par
The debate over the relevance of density dependence has been at times acrimonious, as summarised in~\cite{turchin95}. The quote from that paper which we take as our starting point on this issue is that available evidence ``is entirely consistent with the universal applicability of the density dependence model.''~\citep[][p. 31]{turchin95}. As such, we seek to make what statistical inferences we can about the magnitude and time period of such effects.\par
There are several biological hypotheses as to the causes of density dependence, both in general and in the specific case of North American ducks, our motivating example. These have differing implications for the likely degree of density dependence to be expected in such species.\par
We analyze ten species of duck, including both diving and dabbling ducks, between which there is reason to expect a distinction in density dependence profile. The hypothesis tested (and to an extent borne out) by~\cite{jamieson04} was that diving ducks might, in response to a poor year (low habitat and/or food availability), delay breeding for a year. This would imply a delayed density dependence in diving ducks that would not be present in dabbling ducks.\par
In contrast,~\cite{sargeant84} looked at red fox (\emph{vulpes vulpes}) predation on both diving and dabbling ducks, and concluded that dabbling ducks are significantly more vulnerable to predation of this kind. The red fox is only one predator of ducks in North America, but it is one of the primary predators and, in common with many other duck predators, it is a generalist. A hypothesis of~\cite{bjornstad95}, tested in~\cite{viljugrein05} suggests that this would induce more immediate density dependence in the affected species, since both ducks and eggs are potential predatory targets. This would imply both first and second order density dependence in dabbling ducks; less so in diving ducks.\par
It is apparent that there are hypotheses that produce differing predictions as to the nature of density dependence in these species. We aim to provide a thorough statistical analysis using historical count data provided by~\cite{USFWS10}. 
We will take a Bayesian standpoint when analyzing these data. This is not because a classical analysis is impossible, but rather because we believe that common sense can be translated into a meaningful, informative parameter prior.\par
Inference about the degree of density dependence under this framework is a Bayesian model selection problem.~\cite{link06} illustrate the principles and some of the issues inherent to this class of problem. We demonstrate that choosing an informative prior (using simple rules which we will describe) is both necessary for a balanced model selection procedure, and improves the accuracy with which we can predict future population levels.\par
The outline for the paper is as follows. First we summarise a widely used model for density dependence in the following subsection. Then in section 2 we consider the problem of choosing a Bayesian prior to use in our analysis. Section 3 is a simulation study to exhibit the improvements we offer over previous approaches, before we analyze real data in section 4. We finish with a discussion of our results and lessons learned that can usefully be applied to a wider class of problems than the specific case of density dependence in North American ducks.
\subsection{An Autoregressive Model for Density Dependence}
We consider density dependence model from~\cite{dennis94}. Let $x_t$ be the log-population size in year $t$. The evolution of $x_t$ over time is governed by the stochastic update
\begin{equation}
\label{dennis-taper}
x_t = x_{t-1} + b_0 + \sum_{i=1}^{k} b_i e^{x_{t-i}} + \epsilon_t,\quad \epsilon_t \sim N(0,\sigma^2).
\end{equation}
The parameters are interpreted as\par
\begin{tabular}{l@{\hspace{0.1in}:\hspace{0.2in}}l}
$k$&degree (maximum time lag) of density dependence, in years.\\
$b_0$&uninhibited exponential growth rate\\
$b_{1:k}$&density dependence effects at different time-lags\\
$\sigma^2$&species (and unmodelled covariate) volatility
\end{tabular}\par
The number of $b$ parameters is $k+1$, so $k$ is a model order parameter. If $k=0$, then this process simplifies to a random walk with drift. Also, if several different mechanisms induce a density dependence effect at the same time lag, then the appropriate component of ${\bf b}$ will in effect be a summary statistic measuring the sum of all effects at that time lag.\par
We do not in general observe a true and accurate count of the species abundance. We observe data $y_t$ which will include noise which may vary in intensity from year to year. We assume that this observation process is Gaussian, i.e.
\begin{equation}
\label{modeleq2}
y_t \sim N(x_t,S_t^2)
\end{equation}
and we assume that $S_t$ is known for each year $t=1,\dots,T$. The full model as specified by equations~\ref{dennis-taper} and~\ref{modeleq2} is thus a state space formulation.
\section{Prior Selection}
To perform a full Bayesian analysis and fit of this model, we need to specify a prior for each parameter that is not directly specified by the model itself.\par
We give a uniform prior to $k$, over $\{0,\dots,5\}$. We believe it is implausible that density dependent effects could operate on a longer timescale than this. In particular, the hypotheses that we wish to assess are only concerned with density dependence up to second order. Our prior gives no preference to one time lag over another in this range, so that we can assess the evidence provided by the data in favour of each model. This is similar to a Bayes Factor, which can be used to compare the fit of different models~\citep{kass95}.\par
An immproper inverse gamma (0,0) prior is assigned to $\sigma^2$. This is mostly for reasons of Bayesian conjugacy --- the rate of learning is high for this parameter and the prior shape makes little difference.\par
The distribution of $x_{1:5}$ might not be specified by the model (depending on $k$ --- if $k=2$ for example, then we need to specify the distribution of $x_1$ and $x_2$, and the model gives us the distribution of $x_{3:5}$). In order to have a consistent likelihood across all models, we consider the observed likelihood function $p(y_{1:5} | x_{1:5})$ as a (density) function of $x_{1:5}$ and treat it as our prior. Naturally we do not count it twice, so it is removed from the likelihood, as well as those systemic terms relating to the evolution of $x_{1:5}$. Thus, for all models, the first model-driven term in the likelihood is $p(x_6 | x_{1:5},\sigma^2,{\bf b})$. The final parameter that requires a prior is ${\bf b}$, but there is a pitfall to be aware of before we make our choice.
\subsubsection*{Lindley's Paradox}
It has been known for some time~\citep{lindley57} that choosing a vague (high-variance) prior for within-model parameters (except for parameters common to all of them, such as $\sigma^2$) will bias the model selection routine in favour of simple models. This is discussed in depth in~\cite{link06}. In the limiting case where an improper flat prior is used, the posterior model probabilities will always be degenerate in favour of the model with the fewest parameters.
Lindley's paradox therefore implies that we cannot take a diffuse Normal prior for ${\bf b}$, since this would lead to selecting $k=0$, even if the data produced a likelihood that was higher for other models (hence the paradox).\par
In light of this it is clear we must choose an informative prior, but the question arises as to how to choose an informative prior when one has, apparently, no information. We now show that an informative choice can be reached just by excluding certain pathological cases that we would not expect to arise in the biological systems in question.
\subsection{Stability Considerations}
The population evolution model is simple to simulate from. When one does so one notices that for certain parameter values, the population fluctuates wildly or grows very rapidly until the computer suffers numerical overflow. However, for other values, the population reaches a stable threshold after a period of time (regardless of its starting value) and then does not move too far from this. We refer to this level as the carrying capacity, since it is the maximum level for which the expected population trajectory is not downwards. We would like to restrict our parameters to values that produce a (finite) carrying capacity (exempt from this is the null model $k=0$, as it can never have a carrying capacity). We will demonstrate that a diffuse independent Normal prior does not always lead to the stable scenario, but there are other priors that do (at least much more often).\par
Consider the deterministic analog of the model equation~\ref{dennis-taper} with no measurement error, and suppose that we observe a string of $k$ years where the population is at a constant level $x_{1:k} = x$. Then $$x_{k+1} = x + b_0 + \sum_{i=1}^k b_ie^x.$$ If $b_0$ and $\sum_{i=1}^k b_i$ are of different sign (and $k$ is at least 1), then we can solve for when $x_{k+1} = x$, and we find that this corresponds to 
\begin{equation}
\label{stable-level}
x = x^* = \log\left(\frac{-b_0}{\sum_{i=1}^k b_i}\right).
\end{equation}
This exposes an inherent asymmetry in this model, that $b_0$ and the sum of the other components of $b$ need to be of different sign to produce stable populations. This is not captured in an independent Normal prior. In addition, it raises the problem of estimating the carrying capacity $x^*$. We are constructing a prior, so ``peeking'' at the data should be avoided where possible. The approach we suggest is to center the observed data (on the log scale) so that the carrying capacity should correspond approximately to $x^* = 0$. Thus if you have data on a well-established and stable species, you should center to the mean across all of the time series, whereas if you are analyzing a population that (say) only achieves a stable level in the last fifteen years of a 50-year study, then it should be centered so that the mean of the last fifteen years is zero on the log-scale. This is equivalent to multiplying the data so that the geometric mean over that time period is 1. Optionally, the carrying capacity could be introduced as another parameter and given a prior, but that is not an approach we consider, because it is difficult to get an independent estimate that might inform such a prior.\par
We have suggested centering data, however the model is not invariant to such a transformation. For large populations, a density dependent effect of a particular magnitude will require a smaller ${\bf b}$ than for smaller populations. This is because the $i$-th density dependent effect is equal to $b_i e^{x_{t-i}}$. If we do not center the data, we must incorporate some measure of the overall magnitude of the data into the prior (as done in~\cite{jamieson04}). If we center, then we do not need to look at the data in order to inform our prior.
% It is worth considering though, that natural populations are typically fairly large (commonly thousands and sometimes millions) and on such scales the sensitivity to $b$ is extreme ($x_t$ is 10 to 20, but $e^{x_t}$ is the original population size). Therefore the likelihood-induced variance of $b$ (apart from $b_0$) would be extremely small. In any case, it is common (\cite{jamieson04},~\cite{viljugrein05}) to take data as reported in thousands or millions anyway (effectively dividing the raw data by the same). Once we have decided to divide the data by a constant, we might as well choose it to be something that makes it easier for us to choose the prior, rather than an arbitrary round number. This also avoids basis conditioning issues, which may make it hard to distinguish between density dependent effects at different lags.\par
If we take the carrying capacity to be $x^* = 0$, then, rearranging~\ref{stable-level}, we get
\begin{equation}
\label{deg_norm}
b_0 = - \sum_{i=1}^k b_i.
\end{equation}
Thus $b_0$ is perfectly negatively correlated with each of the other components of $b$. If we take an independent Normal $(0,\sigma_b^2)$ prior for $b_{1:k}$ then this suggests that the joint prior for $b_{0:k}$ should be the degenerate Normal
\begin{equation}
\label{b-prior1}
b_{0:k} \sim N\left({\bf 0},\left(
\begin{array}{cccc}
k \sigma_b^2 & -1 & \dots & -1\\
-1 & \sigma_b^2 & 0 & 0\\
\vdots & 0 & \ddots & 0\\
-1 & 0 & 0 & \sigma_b^2
\end{array}
\right)\right).
\end{equation}
This is degenerate in the sense that the covariance matrix does not have full rank, and only those values of ${\bf b}$ for which~\ref{deg_norm} holds will have nonzero likelihood. In practice this only applies to the deterministic model, and a small amount $h$ would be added to the variance of $b_0$ to allow for mis-estimation of $x^*$. This is because there will always be probabilistic drift towards the carrying capacity, and by allowing some additional variation in $b_0$, we introduce the requisite additional flexibility into the model. The choice of $h$ also dictates the prior under the null $k=0$ model, so a reasonable value might be obtained by considering the variance of symmetric Gaussian random walks over time. For example, a value of $h=0.04225$ corresponds to a process that is as likely as not to at least halve or double in five years. In other words, if $\;Z \sim N(0, 5\times 0.04225)\;$ then $\;{\mathbb P}(Z \in [-\log(2),\log(2)]) = 1/2$. This is the value we use in all of our priors which have $h$ as a parameter.\par
We now consider the effect of small perturbations about carrying capacity. We will see that this restricts even further the set of parameter values that yield a dynamical system we might expect to see in a natural population.\par
\begin{figure}[!ht]
\includegraphics[width=\textwidth]{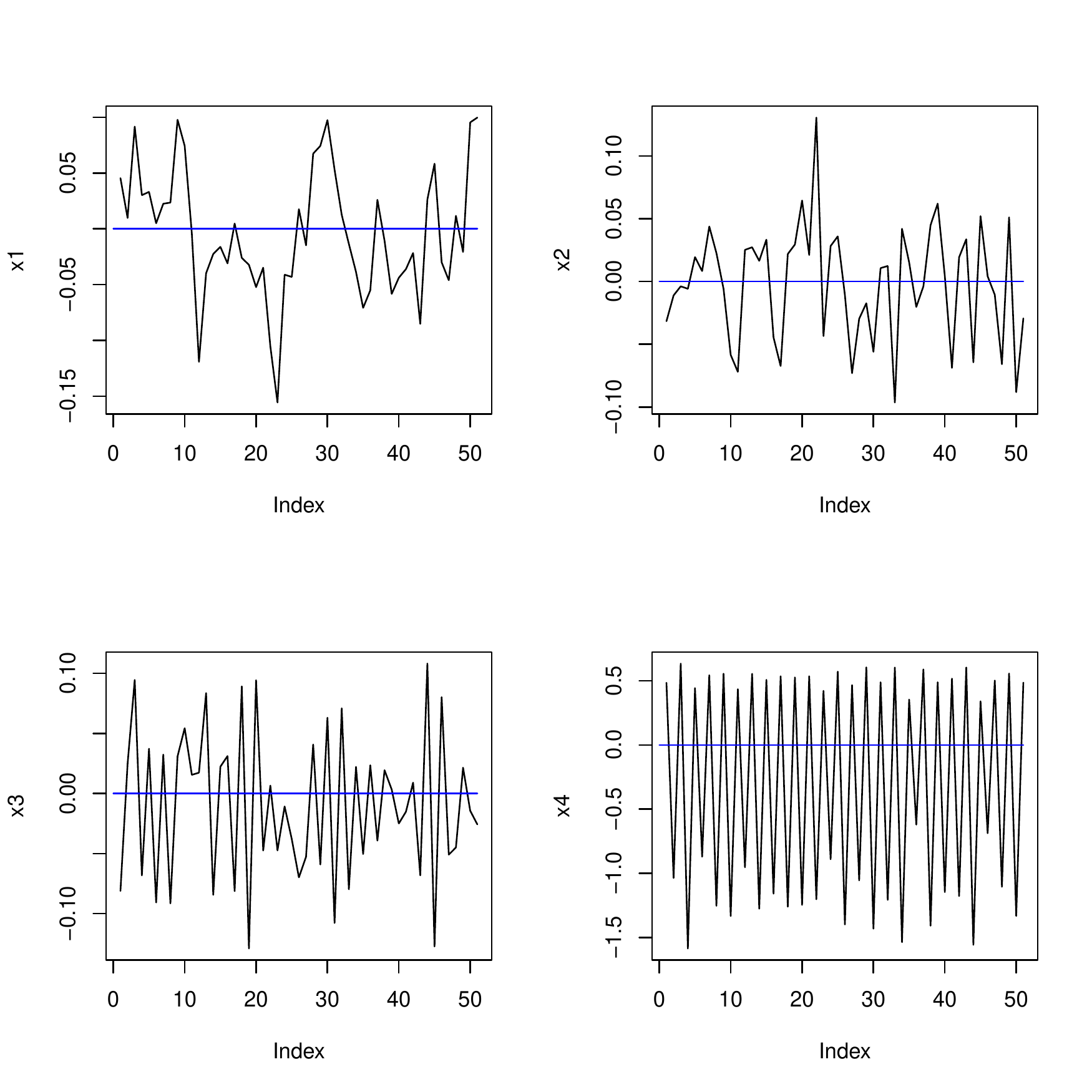}
\caption{Simulations from the autoregressive model, with ${\bf b} = (1/2,-1/2)$, $(1, -1)$, $(3/2, -3/2)$ and $(5/2, -5/2)$. Note that the last of these is a stable exception to the usually unstable case $b_0 > 2$. $\sigma = 0.05$ for all of these, with the process driving the greatly increased variance for the last simulation. There is no measurement error, and we observe from $t=100$ to $t=150$ starting at $(x_1,x_2)=(0,0)$.}
\label{dt_sims1}
\end{figure}

Suppose that $\;x_{1:k} = (0,\dots,0,\delta)$. Then we may be in one of several scenarios (equation~\ref{stable-level} is assumed to hold):
\begin{enumerate}[(a)]
\item $\sum_{i=1}^k b_i$ is positive. In this case, regardless of the sign of $\delta$, the population is unstable and will diverge from 0. Carrying capacity is undefined.
\item $-1 < \sum_{i=1}^k b_i < 0.$ The population returns monotonically towards capacity.
\item $-2 < \sum_{i=1}^k b_i < -1.$ The population oscillates around capacity, with decreasing magnitude.
\item $\sum_{i=1}^k b_i < -2.$ The population oscillates around zero, but usually with much greater magnitude than above. If all of $b_{1:k}$ are negative, then the oscillations will quickly reach a consistent (perhaps large) magnitude, but if any of $b_{1:k}$ are positive, then the population is probabilistically unbounded i.e. with probability 1, as $t \rightarrow \infty$, $x_t \rightarrow \infty$. In the latter case, capacity is again undefined.
\end{enumerate}
Plots of simulated population trajectories for all four cases are given in figure~\ref{dt_sims1}.
We contend that the second of these is most likely to be characteristic of a natural population, but that perhaps some allowance might be made for the third. The first and fourth are considered unlikely to arise in the natural world.\par
This means that had we chosen a prior of the form~\ref{b-prior1} then we would unintentionally be making a strong prior assumption about the model order. For example, if $k=1$, then $\sum_{i=1}^k b_i$ is a $N(0,\sigma^2_b)$ random variable, with a corresponding probability of lying in $[-2,0]$. If $k=2$, then $\sum_{i=1}^k b_i$ has a $N(0,2\sigma^2_b)$ distribution, with correspondingly reduced probability of lying in this interval. This could be thought of as a manifestation of Lindley's paradox. If for example $\sigma^2_b = 1$, then the chance of being in the prior-plausible region under $k=1$ would be $48\%$. Under $k=5$, that chance shrinks to $31\%$. The difference is even more pronounced if $\sigma^2_b$ is higher. Thus, we would be accidentally favouring simple models.\par
A logical refinement of~\ref{b-prior1} is to keep the distribution of $\sum_{i=1}^k b_i$ constant, and to restrict to cases where $-2 < \sum_{i=1}^k b_i < 0.$ This is
\begin{equation}
\label{b-prior2}
b_{0:k} \sim N\left({\bf 0},\left(
\begin{array}{cccc}
\sigma_b^2+h & -1 & \dots & -1\\
-1 & \sigma_b^2/k & 0 & 0\\
\vdots & 0 & \ddots & 0\\
-1 & 0 & 0 & \sigma_b^2/k
\end{array}
\right)\right)
\end{equation}
restricted to the aforementioned set. This is easy and quick to sample from by rejection sampling.
%\begin{enumerate}{}{}
%\item Sample $b$ from the above multivariate normal.
%\item If $-2 < \sum_{i=1}^k b_i < 0$ then keep the sample and terminate.
%\item If $0 < \sum_{i=1}^k b_i < 2$ then take $-b$ as the sample and terminate.
%\item If both of the above steps failed, start again.
%\end{enumerate}
%The acceptance rate of this algorithm depends on $\sigma^2_b$ and is equal to $p(Z \in [-2,2])$ where $Z \sim N(0,\sigma^2_b)$. For example when $\sigma^2_b = 1$ this acceptance rate is a little over 95\%.
%If $\sigma^2_b \leq 1$ then this will have an acceptance rate of at least 95\%.\par
This prior also has the attractive property that the marginal distribution of $b_0$ is the same under all models except $k=0$, so we are equally willing to entertain density dependence effects at different time-lags and we have not unintentionally biased our model prior towards small $k$, since the prior probability of a model where carrying capacity is defined is the same for all $k>0$.\par
\subsection{Shrinkage}
The principle of shrinkage derives from the classical problem of estimating the mean of a multivariate Normal distribution, subject to assumptions about its variance. It can be shown~\citep{stein55} that simply taking the sample mean is inadmissible, provided the dimension is at least three. In other words, a shrunk estimate will provide better (in terms of mean square error) predictions of future observations drawn from the same distribution. We use this idea to motivate an alternative choice of prior, which will have an artificially reduced variance.\par
It must be noted that the improved predictive power shrinkage allows is at the cost of bias. Such bias-variance tradeoffs are common in model selection problems.
\section{Analysis of Simulated Data}
Before we look at observed abundance data, we analyze some simulations of populations which follow the specified dynamics. We have two simulated datasets with the parameters (1) $k=1,\;{\bf b} = (0.5,-0.5)$ and (2) $k=2,\;{\bf b} = (0.5, -0.1, -0.4)$. Both simulations share the parameters $\sigma = 0.05 = S_t {\text{ for each }} t.$ Both series have 501 years of data (this is considerably longer than the real survey, so we can see how much we can expect to learn about the model parameters in the future). We consider five prior choices for ${\bf b}$:
\begin{enumerate}{}{}
\item Independent Normal, variance 5 (primarily as an illustration of Lindley's Paradox).
\item Independent Normal, variance 1 (a baseline for comparison).
\item Multivariate Normal with covariance matrix from the modified version of~\ref{b-prior1}, and $\sigma^2_b = 1, h=0.04225$.
\item A shrinkage-inspired prior: Normal with covariance matrix based on~\ref{b-prior2}:
\begin{equation}
b_{0:k} \sim N\left({\bf 0},\left(
\begin{array}{ccccc}
\sigma_b^2+h & -\sigma_b^2/k & \dots & \dots & -\sigma_b^2/k\\
-\sigma_b^2/k & \sigma_b^2/k & 0 & 0 & 0\\
\vdots & 0 & \sigma_b^2/k & 0 & 0\\
\vdots & 0 & 0 & \ddots & 0\\
-\sigma_b^2/k & 0 & 0 & 0 & \sigma_b^2/k
\end{array}
\right)\right)
\end{equation}
and again $\sigma^2_b=1,h=0.04225$.
\item As (4), but with smaller variance ascribed to later components of ${\bf b}$:
\begin{equation}
\label{shrinkage2}
b_{0:k} \sim N\left({\bf 0},\left(
\begin{array}{ccccc}
\sigma_b^2+h & -\sigma_b^2*d & \dots & \dots & -\sigma_b^2*d/k\\
-\sigma_b^2*d & \sigma_b^2*d & 0 & 0 & 0\\
\vdots & 0 & \sigma_b^2*d/2 & 0 & 0\\
\vdots & 0 & 0 & \ddots & 0\\
- \sigma_b^2*d/k& 0 & 0 & 0 & \sigma_b^2*d/k
\end{array}
\right)\right)
\end{equation}
$d$ is suitably defined so that the sum of variances of $b_{1:k}$ is $\sigma^2_b$. In fact under this restriction$$d = \frac{1}{\sum_{j=1}^{k} 1/j}$$ in equation~\ref{shrinkage2} for $k \geq 1$. Notice that both priors (4) and (5) have the same total variance for ${\bf b}$, as long as $k > 0$. This is deliberate, as discussed earlier.
\end{enumerate}
The choice between the last two priors largely depends on whether one considers the assumption that longer lags tend to be smaller in size to be suitable {\emph{a priori}}. We will see that they do not provide substantially different estimates or predictions, but then we only consider simulations for low values of $k$.\par
We use a Particle Learning method~\citep{carvalho10} combined with Reversible Jump MCMC~\citep{green95} to produce a sample from the posterior for each simulation.% The algorithm can be summarised thus:
%\begin{enumerate}{}{}
%\item Produce an initial sample of N ``particles'' $P_{1:N}$, each of which consists of a value for $k$, ${\bf b}$, $%\sigma$ and $x_{1:k_{\max}}$ drawn from their respective priors.
%\item To update from time $t$ to time $t+1$, first calculate for each particle its predictive accuracy$$w^{(t+1)} = %\int L(y_{t+1} | x_{t+1},S_{t+1}) p(x_{t+1} | x_{1:t},k,{\bf b}) dx_{t+1}.$$ This behaves as the incremental weight for that particle.
%\item Resample the particles with replacement, sampling each with probability proportional to $w^{(t+1)}$ for that particle.
%\item For each particle, sample a value of $x_{t+1}$ from $p(x_{t+1} | x_t, k, {\bf b})$.
%\item Update the values of $k$ and ${\bf b}$ for each particle by sampling from the equilibrium distribution of the update detailed in~\cite{godsill97}.
%\end{enumerate}
This produces a weighted sample from the posterior distribution of models, parameters and hidden states. We are also able to chart the posterior as it evolves over time, as more data are added.
\subsection{Model Selection Results}
\begin{figure}[ht]
\centering
\subfigure[Independent prior, variance 5]{\includegraphics[width=5.5cm]{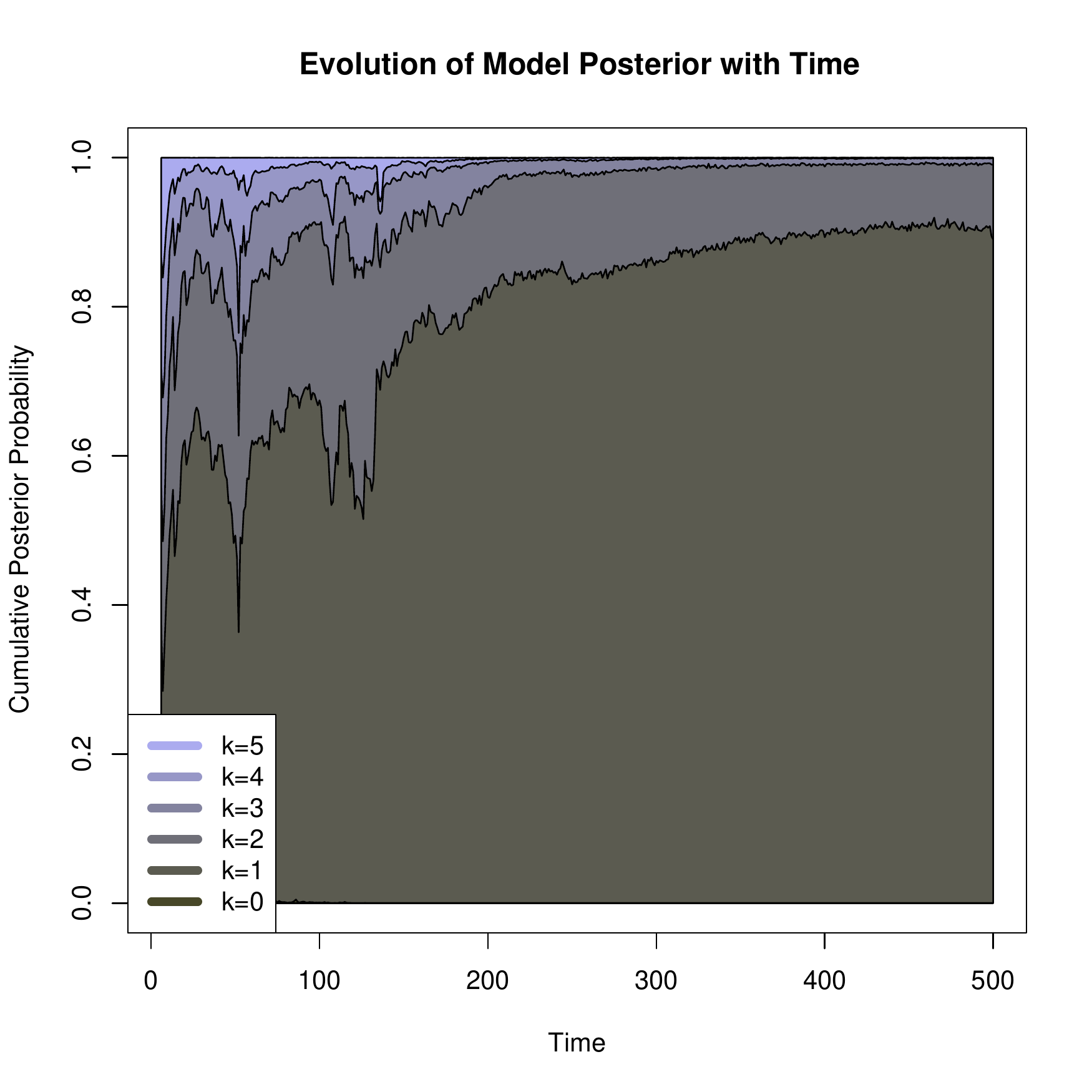}}\hspace{1in}
\subfigure[Independent prior, variance 1]{\includegraphics[width=5.5cm]{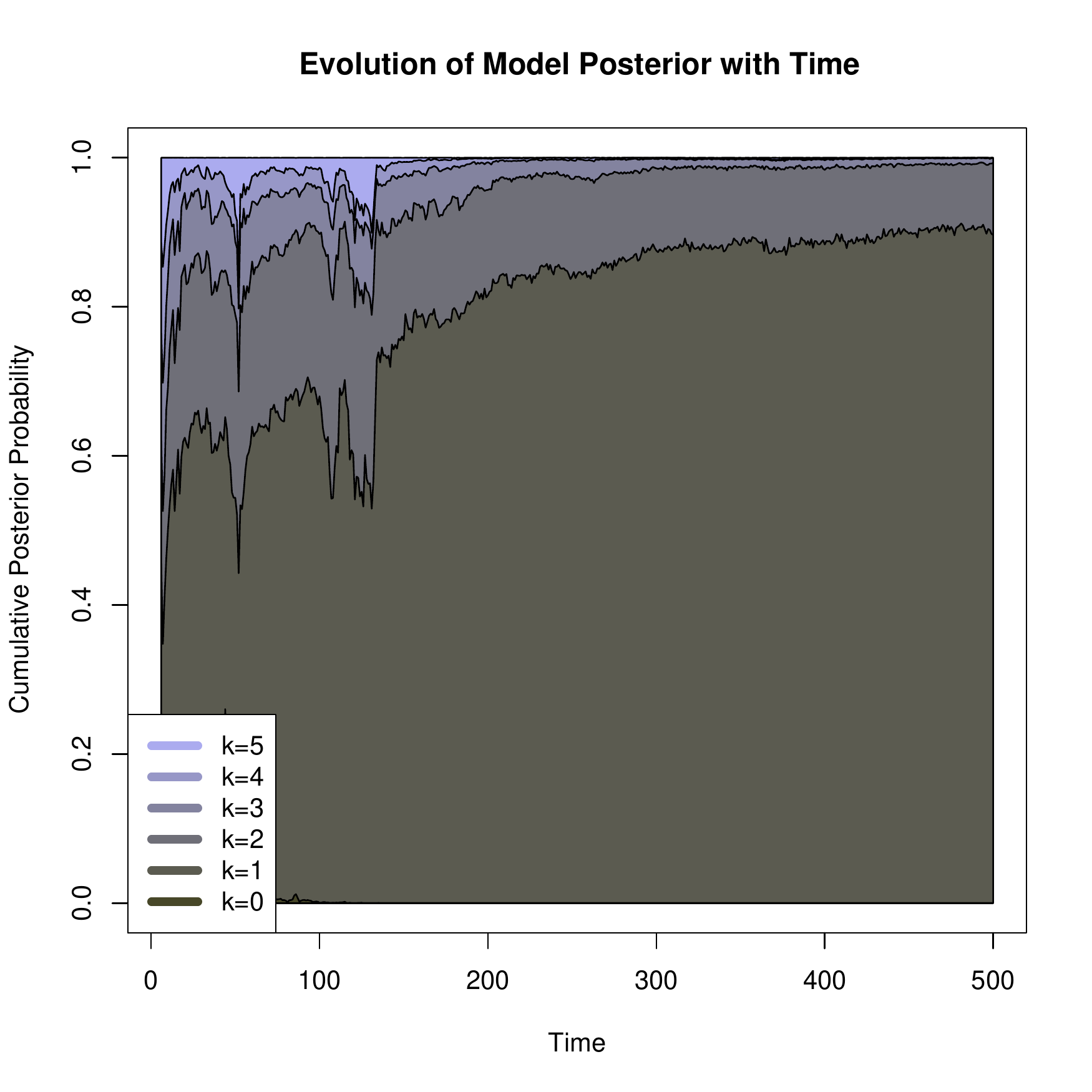}}\\
\subfigure[Correlated prior, variance 1]{\includegraphics[width=5.5cm]{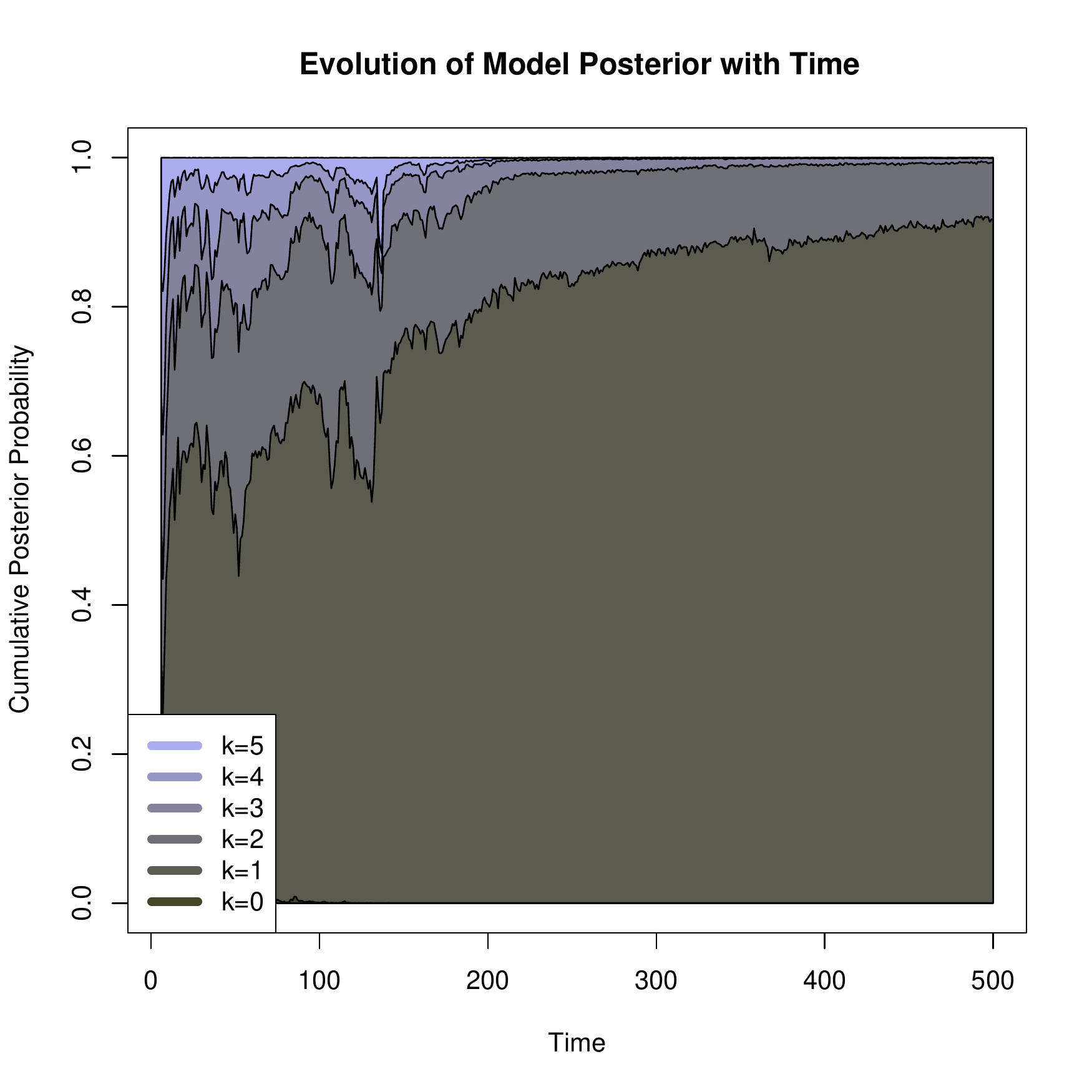}}\hspace{1in}
\subfigure[Shrinkage prior, variance $1/k$]{\includegraphics[width=5.5cm]{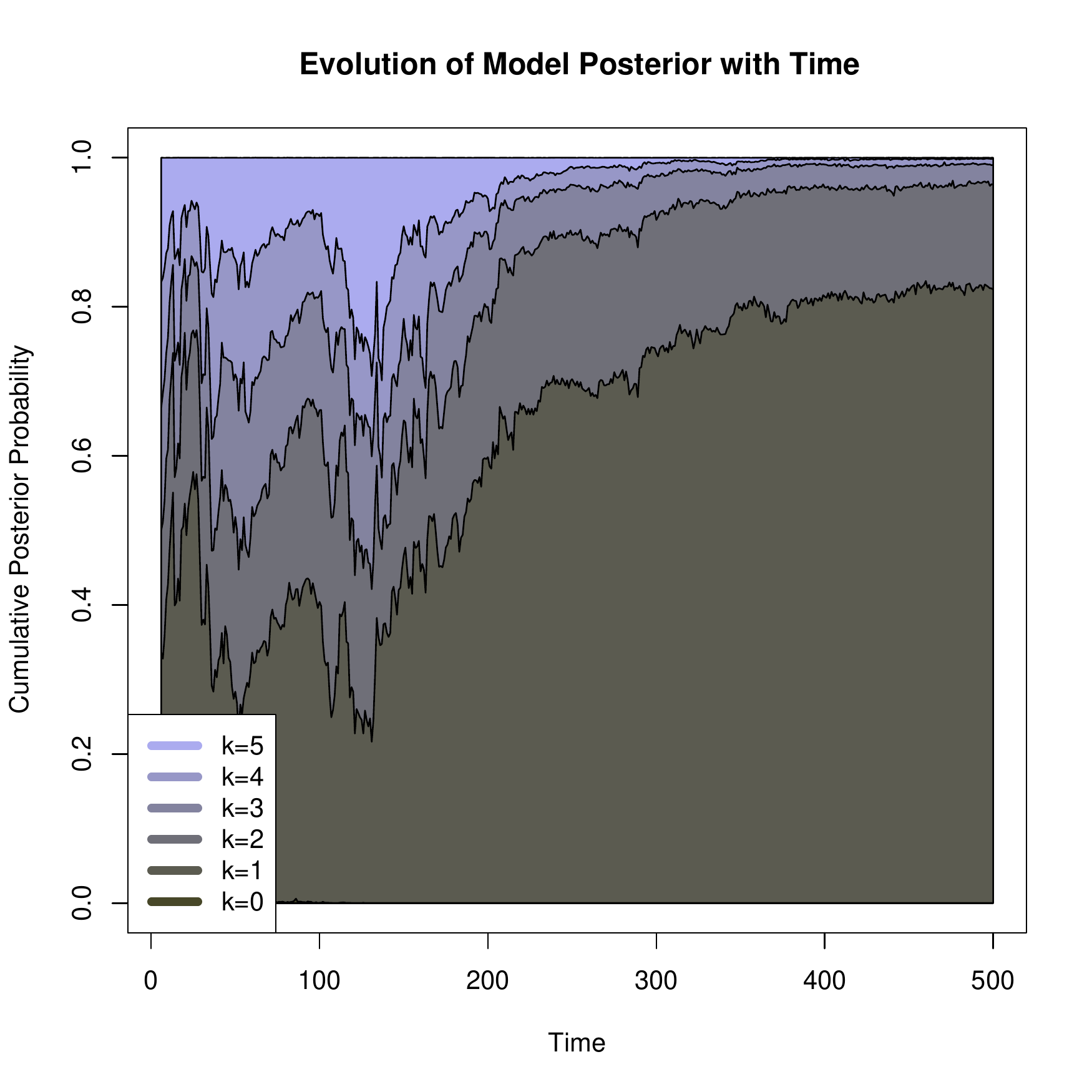}}
\caption{Evolution of the posterior model distribution over a long time span when $k=1$, for four different prior choices (1)--(4). The fifth posterior is almost identical to the fourth.}
\label{evolving_posterior_long1}
\end{figure}

The evolution of the posterior for $k$ in the $k=1$ simulation is shown in figure~\ref{evolving_posterior_long1}.
The results for the second simulation are not pictured; they are qualitatively similar (except that the majority of the posterior mass is on $k=2$ instead of $k=1$, after a similar period of time). This figure makes it clear that while we can learn the value of $k$, we can only do so slowly and given that we only have fifty or so years of duck abundance data, we cannot expect a conclusive model selection posterior. This makes it particularly important that we choose a parameter prior that will not influence the model selection process, since the signal from the data is quite weak.\par
Another point of interest is the posterior at $t=6$, i.e. after only one residual is taken into account. This is the first point at which we have a model posterior, and we can see for the independent priors that this posterior is far from uniform. This is one quantification of the model selection bias induced by the independent priors. 
\section{Analysis of Observed Data}
In total, eleven species are analyzed. Seven of these are dabbling ducks: Mallard (\emph{Anas platyrhynchos}), American Wigeon (\emph{Anas americana}), Gadwall (\emph{Anas strepera}), Green-Winged Teal (\emph{Anas crecca}), Blue-Winged Teal (\emph{Anas discors}), Northern Shoveler (\emph{Anas clypeata}) and Northern Pintail (\emph{Anas acuta}).
The remaining four are diving ducks, two of which are amalgamated: Redhead (\emph{Aythya americana}), Canvasback (\emph{Aythya valisineria}) and Greater and Lesser Scaup (\emph{Aythya marila} and \emph{Aythya affinis}).\par
The data, as supplied by~\cite{USFWS10}, include both an estimated annual count and an estimate of the observation error. We treat the observation error as exact. The posterior model probabilities for each species, using the shrinkage prior (4), are summarised in table~\ref{pmp_realdata}.\par
\begin{figure}[ht]
\begin{center}
\begin{tabular}{|l|cccccc|}
\hline
Species&$k=0$&$k=1$&$k=2$&$k=3$&$k=4$&$k=5$\\
\hline
Mallard&0.1718&0.1882&0.208&0.0783&0.2399&0.1138\\
A.Wigeon&0.0238&0.4192&0.2637&0.1337&0.0601&0.0995\\
Gadwall&0.6805&0.1664&0.0553&0.0191&0.045&0.0338\\
G.W.Teal&0.6816&0.0982&0.052&0.053&0.0588&0.0563\\
B.W.Teal&0.4422&0.3197&0.1347&0.0582&0.0276&0.0176\\
N.Shoveler&0.4906&0.0756&0.2494&0.0942&0.0421&0.0481\\
N.Pintail&0.2733&0.2319&0.2707&0.1057&.0322&0.0862\\
\hline
Redhead&0.3239&0.0671&0.2005&0.1489&0.1355&0.124\\
Canvasback&0.0299&0.5284&0.192&0.0939&0.0922&0.0636\\
Scaup&0.5764&0.1454&0.1349&0.0684&0.0418&0.0331\\
\hline
\end{tabular}
\caption{Posterior model probabiliites for each duck species, using a shrinkage prior.}
\label{pmp_realdata}
\end{center}
\end{figure}
None of the posteriors are conclusive as to the order of density dependence. We expect this from the simulation study; even with data that we \emph{know} follows a particular instance of the model, we can only expect perhaps a 60\% posterior probability for that model after this length of time. It would be optimistic to expect the same level of agreement with real data.
\subsection{Predictive Accuracy}
\begin{figure}[ht]
\centering
\includegraphics[width=9.5cm]{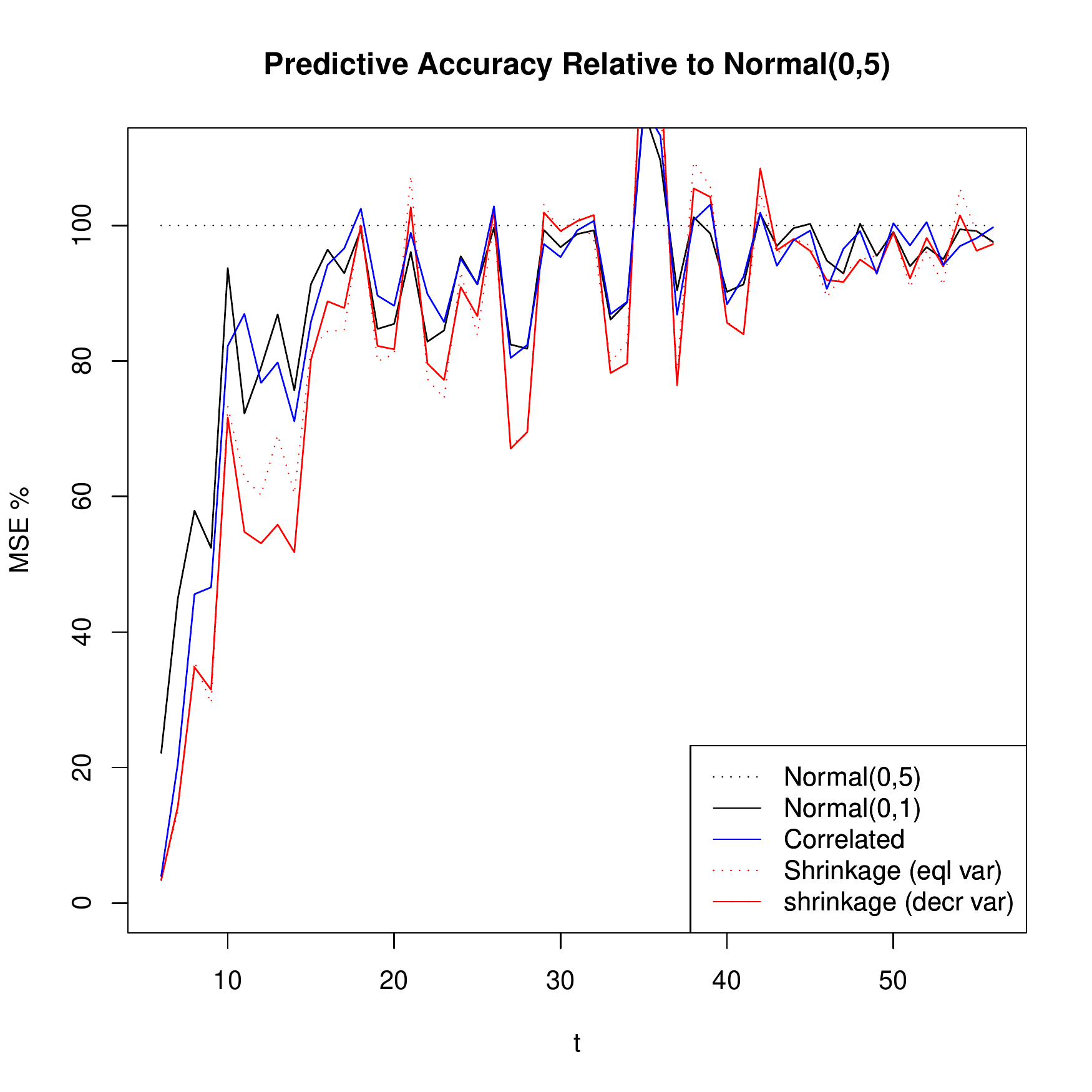}
\caption{Mean squared predictive error for different priors, scaled so that the Normal(0,5) prior is at 100\%. American Wigeon data.}
\label{MSEvs5}
\end{figure}

It is impossible to assess the quality of the $k$ posterior, since we have nothing with which to compare it. We can however look at the ability of the posterior at a given time point to make predictions of future numbers.  These can then be compared with our best guess of the truth for that year (which the predictions were made without knowledge of.) A simple quantity that measures predictive accuracy is the one step ahead Mean Square Error $\text{MSE}(t) = {\mathbb E}(\hat{x}_t - \tilde{x}_t)^2$ where $\hat{x}_t$ is the prediction of $x_t$ from the particle set at time $t-1$ and $\tilde{x}_t$ is the ``smoothed'' state estimated from the particle set at time $T$. We seek to minimize MSE. As a typical example of the relative performance of each prior, figure~\ref{MSEvs5} shows the evolution of the MSE over time, using the $N(0,5)$ prior as a baseline, for the American Wigeon data.
After a certain time, the MSE becomes approximately equal for all priors. This shows that the data has overwhelmed the prior in terms of information. Before then, there is significant disparity in MSE for the different priors, and while the correlated prior offers a mild improvement over the independent one, the shrinkage priors clearly outperform the others for up to 30 years.\par
The MSE can sometimes be slightly misleading, since predictions are correlated (as are the quantities they are predicting). One measure to correct this is the Mahalanobis distance~\citep{mahalanobis36}. This is based on taking a Gaussian approximation to the predictive distribution and calculating the expected total squared error over the whole time series. It is given by
\begin{equation}
\label{mahal_dist}
D_M = (\hat{\bf x} - \tilde{\bf x})^T S^{-1} (\hat{\bf x} - \tilde{\bf x}).
\end{equation}
The Mahalanobis distance is not a function of time, it measures performance from start to finish. A low Mahalanobis distance is indicative of good overall predictive accuracy. When we calculate the Mahalanobis distance under each choice of prior for each species, we obtain table~\ref{MD_table}.
\begin{figure}[h]
\centering
\begin{tabular}{|l|ccccc|}
\hline
Species & N(0,5) & N(0,1) & Corr. & Shrink. 1 & Shrink.2\\
\hline
Mallard&20726&5416&1767&1622&1575\\
A.Wigeon&4803&1616&949&818&807\\
Gadwall&2884&1498&1302&1044&1029\\
GW.Teal&4266&2171&1717&1326&1396\\
BW.Teal&7553&2286&1088&1035&992\\
N.Shoveler&4384&1789&1268&1122&1130\\
N.Pintail&19561&5228&2367&1798&1763\\
\hline
Redhead&3361&1636&1117&1005&981\\
Canvasback&3118&1073&529&479&481\\
Scaup&16153&3702&1037&972&879\\
\hline
\end{tabular}
\caption{Mahalanobis Error for different prior choices.}
\label{MD_table}
\end{figure}
The story is broadly the same for all the species, as follows: The independent priors have much more predictive error than the correlated ones (the high-variance prior being worst). The shrinkage priors, as expected, offer improvements over all the others, however there is little difference in accuracy between the two types of shrinkage prior.
\subsection{Interpretation of Results}
We see that for most species, we cannot discount the possibility that $k=0$. This can be interpreted in a few different ways. The simplest explanation (which is also the least likely to be true in the authors' opinion) is that the species do not show density dependent dynamics. It is also possible that the species are in fact far from carrying capacity, so that the density dependent effects are too small to be measured. In that case a hypothesis must be made as to what is keeping the species from reaching capacity, and that is beyond the scope of this study. It might be possible that the numerical nature of the density dependence cannot be projected onto this class of models. If we were to observe the species over a longer time period, or where it were closer to capacity, these differences would likely present themselves in the form of evidence for $k>0$ in the posterior.\par
It is interesting to note that the Mallard (which is the only species for which the posterior-preferred model is greater than $k=2$) is also the species with the highest population count. This is potentially indicative that intra-species competition is a major factor in dabbling ducks, as there is a strong negative correlation between total count and posterior probability that $k=0$ for dabbling ducks. This correlation also could be taken as evidence of the generalist predator hypothesis, which would argue that changes in duck recruitment (i.e. changes to $b_0$) would be met with immediate responses from the predator (so that in fact $b_0$ might change from year to year, but in a way that is probabilistically equivalent to the $k=0$ model with the variance being added to $\sigma^2$ instead).\par
The picture is somewhat different for diving ducks. The aforementioned correlation between raw numbers and apparent density dependence is not apparent here. Again, this is consistent with the generalist predator hypothesis which, taken in conjunction with the reports from \cite{sargeant84} about diving ducks being much less vulnerable to this kind of predation, would suggest a different density dependent structure from that of dabbling ducks. Even here though, there is still appreciable posterior probability that $k=0$ in two out of three cases.\par
The hypothesis of~\cite{jamieson04}, that diving ducks were in general more density dependent than dabbling ducks, is not really borne out by this analysis. The authors of that paper used independent priors with different variance for each species. As one example, for the Blue Winged Teal, the authors had an independent prior variance of 3, and came to a posterior that was 73\% in favour of $k=0$, and almost all the rest of the mass was for $k=1$. We have demonstrated that this is largely an artifact of Lindley's Paradox and our posterior is much less conclusive.\par
\section{Discussion}
We hope that we have demonstrated the importance of a considered choice of prior. A default choice is rarely safe in model selection problems, and we have shown how, by considering whether the carrying capacity is well-defined and trying to exclude cases where it isn't, we can arrive at an informative prior without peeking at the data.\par
A more general principle is that of excluding so-called `unphysical' possibilities from the prior, that is, not allowing parameters to take values which would produce behaviour we know does not happen. We excluded models which did not give rise to a well-defined carrying capacity; the precise nature of the prior restrictions will vary from problem to problem.\par
It is important to consider how a parameter's prior varies between models: a parameter with a different interpretation in different models may well require a different prior in each case. In our example $b_0$ typically had a prior that was different under the null model $k=0$ than in more complex cases. This mirrored the fact that in the null model $b_0$ was interpreted as an overall drift, whereas otherwise it was the counterbalance to the density dependence effects.\par
When we excercise such caution in choosing our parameter priors, we are in a position to judge much more effectively whether the data provide evidence in favour of our hypotheses or not. 
\subsection*{Acknowledgements}
JDL is funded by an Engineering and Physical Research Council Grant (EPSRC) number (to follow). This work was also partially funded by EPSRC grant EP/D06570 4/1 to RBG. Most of this research was conducted while RBG was a Lecturer at the Statistical Laboratory, University of Cambridge.
\bibliography{ducks}
\bibliographystyle{jasa}
\end{spacing}
\end{document}